\documentclass[letterpaper]{jpconf}
\usepackage{graphicx}
\begin{document}
\title{Semileptonic Decays: The Heavy Daughter Quark Limit}

\author{Matthew Dowling}

\address{Department of Physics, University of Alberta, Edmonton, Alberta, Canada T6G 2G7}

\begin{abstract}
We have calculated the rate of the decay $b\to c\ell\overline{\nu}_{\ell}$ to second order in $\alpha_s$ in the limit that the $b$ and $c$ quarks have equal masses.
The results here confirm recent calculations done in the opposite limit where the $c$-quark is much lighter than the $b$-quark.
\end{abstract}

\section{Introduction}
In the effort to find physics beyond the Standard Model, one of the methods being implemented is over-constraining the unitarity triangles that arise from the Cabbibo-Kobayashi-Maskawa (CKM) quark mixing matrix.
One important quantity that must be measured for this is the value of the CKM parameter $|V_{cb}|$.
This is typically done in one of two ways: inclusive or exclusive measurements.
In exclusive measurements specific decay rates, for example $\overline{B}^0 \to D^{*+}\ell^-\overline{\nu}_{\ell}$, are measured and compared with theoretical predictions to extract a value for $|V_{cb}|$.
The inclusive measurements, however, look at all decays from a meson with a $b$-quark to a daughter meson containing a $c$-quark.
Currently, the most accurate measurement we have for $|V_{cb}|$ is from the inclusive decays.
The agreement between the two measurements, however, is only about $2.5\%$.
As such, one would like to improve on these measurements to aid in over-constraining the unitarity triangles.

Looking specifically at the inclusive decays, the current value of $|V_{cb}|$ is,
\begin{equation}
|V_{cb}| = (41.6\pm 0.6) \times 10^{-3},
\end{equation}
with a large portion of the error coming from higher orders in the perturbative calculations \cite{Amsler:2008zzb}.
This determination of $|V_{cb}|$ included the leading and next-to-leading order perturbative calculations along with an order of magnitude estimate of the second order corrections called the Brodsky-Lepage-Mackenzie (BLM) correction.

Naturally, the next step is to calculate the full second order corrections.
This was done in two separate calculations: one done numerically \cite{Melnikov:2008qs} and one analytically \cite{Pak:2008qt}.
The analytic calculation involved an expansion in the $b$ to $c$ quark-mass ratio, $\rho=m_c/m_b$, with a light $c$-quark.
This expansion gave results that were consistent with the numerical results of \cite{Melnikov:2008qs} when the ratio of masses was in the region of the physical value.
If one wanted to go above this value though, the analytic result started to diverge.
This is solely due to the truncation of the expansion in $\rho$ that was done.
In order to extend our knowledge of the second order corrections to $\rho=1$ we have carried out the same calculation but in the opposite limit of the mass ratio i.e. $\rho \approx 1$ \cite{Dowling:2008mc}.
This expansion provides an independent check of the results in \cite{Pak:2008qt} as well as providing accurate values for the decay rate through the full range of the quark-mass ratio.

\section{Methods}
\begin{figure}[t!]
\centering
\includegraphics[width=1\textwidth]{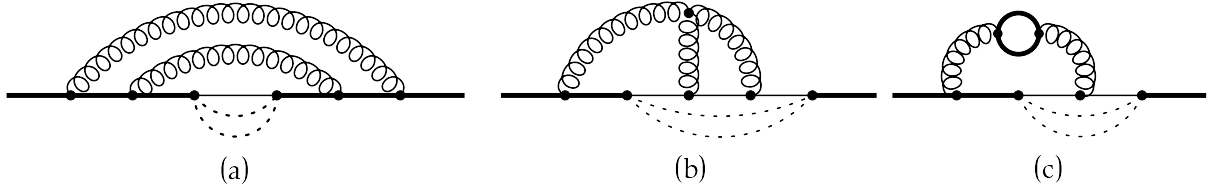}
\caption{\label{fig:diagrams} A sample of diagrams that needed to be calculated for the second order corrections to the decay width. Dashed lines denote leptons and curly lines denote gluons, while thick and thin lines denote $b$ and $c$ quarks, respectively.}
\end{figure}

To calculate the second order corrections to the decay width of $b \to c\ell\overline{\nu}_{\ell}$, we treat the masses of the $c$ and $b$ quarks to be almost equal and expand in small parameter $\delta = 1-\rho$.
The second order corrections involve calculating decays with up to five daughter particles.
We found it convenient to use the optical theorem and instead calculate the imaginary parts of four loop diagrams, some of which are shown in Figure~\ref{fig:diagrams}.
One of the benefits of calculating loop diagrams is that we can then use all of the machinery that has been developed.
In particular, for this problem, the method of asymptotic expansion simplifies the calculation immensely \cite{Smirnov:2002ch,Tkachov:1994pz}.

Asymptotic expansion is a method of separating the large and small momentum contributions, or regions, of the loops in a diagram so that the resulting integrals are more tractable.
In this calculation, we happen to be in a kinematic configuration that was ideal for asymptotic expansion.
When applied, the region with all loop momenta large ends up not contributing to the decay rate.
This region typically gives the most difficult integrals and so our calculation was greatly simplified.
Another benefit of asymptotic expansion in this calculation was that all diagrams could be reduced to five different integrals with at most two loops.

The other important property of loop integrals used, is that of recurrence relations.
This is a method based on integration-by-parts identities \cite{Chetyrkin:1981qh,Tkachov:1981wb} that gives relations between Feynman integrals and it is often the case that the resulting integrals are easier to calculate than the original one being considered.
Specifically, we used the Laporta algorithm \cite{Laporta:1996mq,Laporta:2001dd} to treat the diagrams that have a three gluon vertex: e.g. Figure~\ref{fig:diagrams}b.
We used two programs, ROWS \cite{Pak:jb} and FIRE \cite{Smirnov:2008iw} in order to do this reduction.

With these simplifications in place, we were able to obtain an expansion that is highly convergent and surprisingly easy to compute, especially compared to the expansion from the opposite limit that was done in \cite{Pak:2008qt}: see also \cite{Pak:2008cp}.

\section{Results}
The total width is customarily written in the form
\begin{equation}
\Gamma = \frac{G_{F}^2|V_{cb}|^2m_b^5}{192\pi^3}\left[ X_0 + \frac{\alpha_s(m_b)}{\pi}C_FX_1+\left( \frac{\alpha_s}{\pi}\right)^2 C_FX_2 + \ldots\right],
\end{equation}
\begin{equation}
X_2 = C_FX_F + C_AX_A + T_F(n_lX_l + X_c + X_b),
\end{equation}
with $G_F$ the Fermi constant, and $m_b$ the pole mass.
We have $C_F=4/3$, $C_A=3$, $T_F=1/2$ in QCD and $n_l=3$ the number of light flavours in our theory.
Full analytic results for $X_0$ and $X_1$ can be found in \cite{Nir:1989}, while analytic expansions for $X_2$ can be found in \cite{Pak:2008qt,Dowling:2008mc}.

\begin{figure}[ht!]
\centering
\includegraphics[width=0.75\textwidth]{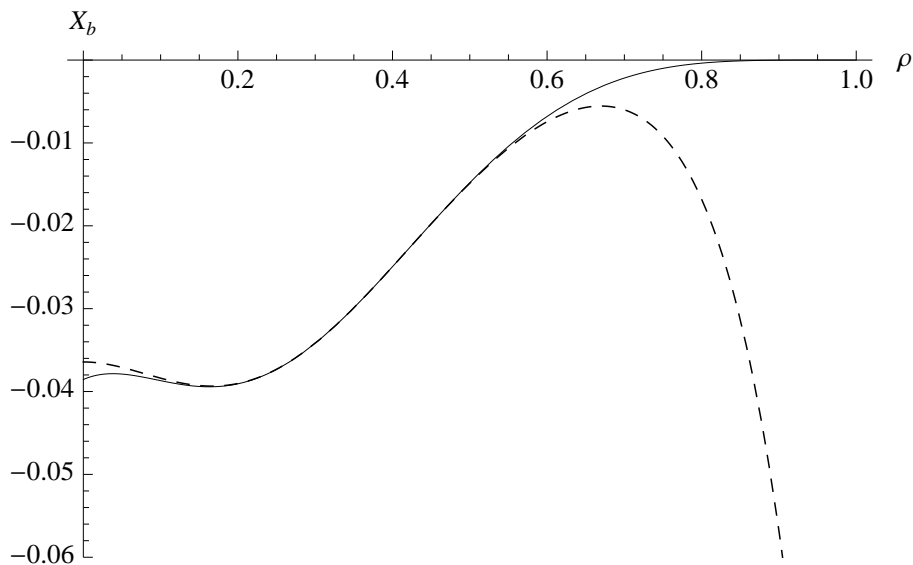}
\caption{\label{fig:xb} Plot of the contribution $X_b$ as a function of $\rho=1-\delta$. The solid line is our expansion, while the dashed line is the expansion from \cite{Pak:2008qt}.}
\end{figure}

The final result of our calculation is an expansion in $\delta$ with $X_c,X_b$ and $X_l$ terms calculated up to $\mathcal{O}(\delta^{15})$, $X_F$ to $\mathcal{O}(\delta^{12})$, and $X_A$ to $\mathcal{O}(\delta^{11})$.
As an example of the results, we provide the $X_b$ term to order $\delta^7$ here.
This term receives contributions from diagrams like that in Figure~\ref{fig:xb}c.
The full results can be found in the source files for the preprint of \cite{Dowling:2008mc}.
\begin{equation}
X_b = \left(\frac{184}{3} - \frac{32}{5}\pi^2\right)\delta^5 + \left(-12+\frac{8}{5}\pi^2\right)\delta^6 +
	\left(\frac{107444}{2835} - \frac{3848}{945}\pi^2\right)\delta^7 + \ldots
\end{equation}
Note that the decay is highly suppressed as it starts at $\delta^5$.
As well, the first two terms include fractions that are simpler than the $\delta^7$ term.
This occurs in all of the second order terms and has an interesting explanation as to why.
These two terms come from the zero recoil limit where $m_c=m_b$ and the mass of the leptons is zero.
This calculation has been done in \cite{Czarnecki:1996gu} and we have verified the first two terms in our expansion using these results.

If we look at the plot of $X_b$ as compared to the expression in \cite{Pak:2008qt}, Figure~\ref{fig:xb}, we can see the $\delta^5$ suppression.
Figure~\ref{fig:xb} also shows how well our expansion agrees with the previous one and we were able to reproduce the minimum at about $\rho = 0.2$.

\begin{figure}[ht!]
\centering
\includegraphics[width=0.75\textwidth]{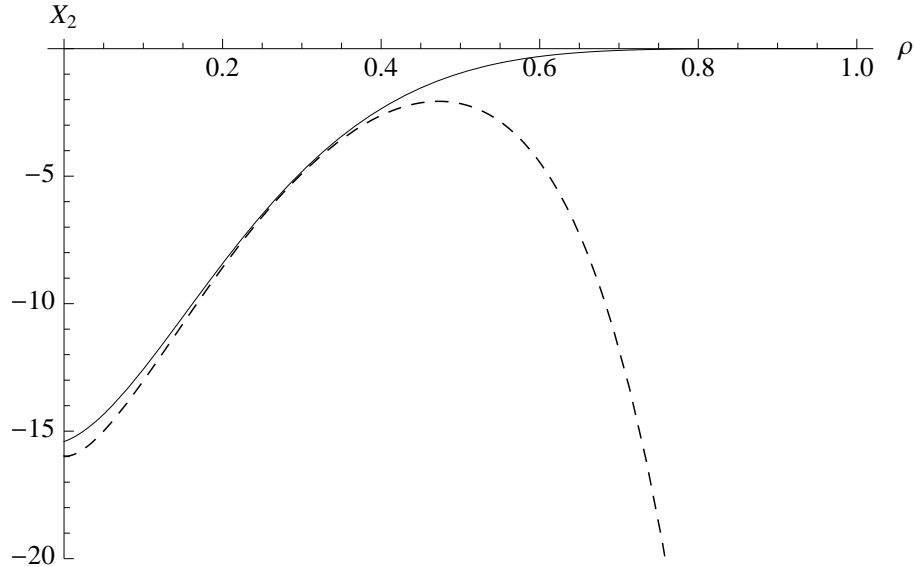}
\caption{\label{fig:full} Plot of the full second order correction as a function of $\rho=1-\delta$. Again, the solid line is our expansion, while the dashed line is the expansion from \cite{Pak:2008qt}.}
\end{figure}

A plot of the full result is shown in Figure~\ref{fig:full}.
In this plot, we can see that the two expansions agree well in the region $\rho \approx 0.1-0.4$.
As well, our expansion is accurate to about $2.5\%$ all the way to $\rho=0,\delta=1$, where our expansion isn't expected to work well.
Given the simplicity of this calculation and this fact, it is conceivable that a calculation of the $\mathcal{O}(\alpha_s^3)$ corrections could be computed with results that extend across the full range of mass ratio $m_c/m_b$.

\section{Conclusions}
We have calculated the second order corrections to the decay $b\to c\ell \overline{\nu}_{\ell}$ in the limit of a heavy daughter quark.
This expansion agrees with a previous expansion in the opposite limit \cite{Pak:2008qt} and verifies some interesting properties of the previous result.
As well, our result is highly convergent and thus gives accurate values for the decay rate at any value of the expansion parameter $\delta$.
Along with the simplification of the calculation through use of procedures like asymptotic expansion and the Laporta algorithm, this gives promise for higher order calculations in the future.

\section*{Acknowledgments}
This work was supported by Science and Engineering Research Canada and the Feynman diagrams were drawn using JAXODRAW \cite{Binosi:2004jx}
\section*{Refrences}
\bibliographystyle{iopart-num}
\bibliography{Database}

\end{document}